\documentclass{iau} 
\usepackage{graphicx}

\usepackage[sort]{natbib}

\usepackage{hyperref} 
\def\aj{AJ}%
\def\actaa{Acta Astron.}%
\def\apj{ApJ}%
\def\aap{A\&A}%
\def\aaps{A\&AS}%
\def\mnras{MNRAS}%
\def\pasp{PASP}%
\def\raa{RAA}%

\title[Accurate photometry with digitized photographic plates of the Moscow collection] 
{Accurate photometry with digitized photographic plates of the Moscow collection}

\author[Sokolovsky, Zubareva, Kolesnikova, Samus, Antipin \& Belinski]   
{K.~V.~Sokolovsky$^{1,2,3}$,
A.~M.~Zubareva$^{4,2}$,
D.~M.~Kolesnikova$^{4}$,
N.~N.~Samus$^{4,2}$,
S.~V.~Antipin$^{2}$,
A.~A.~Belinski$^{2}$}

\affiliation{$^{1}$IAASARS, National Observatory of Athens, 15236 Penteli, Greece \\[\affilskip]
$^2$Sternberg Astronomical Inst. MSU, Universitetskii~pr. 13, 119992 Moscow, Russia \\[\affilskip]
$^3$Astro Space Center, LPI RAS, Profsoyuznaya Str. 84/32, 117997 Moscow, Russia \\[\affilskip]
$^4$Institute of Astronomy RAS, Pyatnitskaya Str. 48, 119017 Moscow, Russia}

\pubyear{2018}
\volume{339}  
\jname{Southern Horizons in Time-Domain Astronomy}
\editors{E. Griffin, R. Seaman, M. Sullivan \& P. Woudt eds.}
\begin{document}

\maketitle

\begin{abstract}
Photographic plate archives contain a wealth of information about positions and brightness celestial objects had decades
ago. Plate digitization is necessary to make this information accessible, but extracting it is a technical challenge.
We develop algorithms used to extract photometry with the accuracy of better than $\sim 0.1$\,m in the magnitude range
$13<B<17$ from photographic images obtained in 1948--1996 with the 40\,cm Sternberg institute's astrograph
($30\times30$\,cm plate size, $10\times10$\,deg field of view) and digitized using a flatbed scanner. 
The extracted photographic lightcurves are used to identify
thousands of new high-amplitude ($>0.2$\,m) variable stars. 
The algorithms are implemented in the free software \texttt{VaST} available at
\url{http://scan.sai.msu.ru/vast/}
\end{abstract}

\firstsection 
\section{Introduction}

Photography was historically the first method to objectively record
brightness of astronomical objects \citep{2001astro.ph..6313K}, however extracting brightness information 
from photographic images was never easy.
While sophisticated  glass plate measuring machines 
including microdensitometers \citep{1984NASCP2317..209S} and iris diaphragm
photometers \citep{1989PASP..101.1038T} were developed, they were generally 
slow and inefficient. Often the most practical way to extract brightness of a
given object from a set of plates was the visual inspection of these plates 
with a magnifying glass. An astronomer would estimate the brightness of 
the object by visually comparing it to the nearby stars of known brightness
\citep{1905PA.....13..453Y}.
An experienced observer can obtain magnitude measurements with 
the accuracy of better than $\sim 0.1$\,m, comparable to that of a measuring machine
\citep{2004JAVSO..32..117D}.

It is now possible 
to digitize the full photographic plate
in a matter of (tens of) minutes in order to make its information content accessible. 
Digitization may serve the additional purpose of preserving the information 
if the original glass plate is lost. 
Observatories around the world make
efforts to digitize their plate collections. 
The DASCH project at Harvard \citep{2009ASPC..410..101G,2012IAUS..285...29G,2013PASP..125..857T} 
as well as Shanghai \citep{2017RAA....17...28Y}, Belgium
\citep{2011MNRAS.415..701R,2012ASPC..461..315D} 
and Tautenburg \citep{2008A&A...477...67H}
observatories use the purposely-built measuring
machines which effectively take digital photographs of plates using a
telecentric lens delivering high image quality and digitization speed.
Observatories including the APPLAUSE collaboration -- Bamberg, Hamburg, Potsdam, and Tartu
\citep{2017AN....338..103W},
Sonneberg \citep{2009chao.conf..311K,2017ApJ...837...85H}, 
Rozhen \citep{2012PASRB..11..201M}, Asiago \citep{2004BaltA..13..665B} and
many others employ commercial flatbed scanners despite their
known drawbacks of being slow and introducing 
the characteristic hacksaw-like pattern of systematic
errors in astrometry \citep{2009ASPC..410..111S}.

In this work we focus on processing a specific series of plates from the Moscow collection
(the ``A'' series) from which we expect the most scientific return after
digitization. This series includes 22300 blue-sensitive $30\times30$\,cm plates
($10\times10$\,deg field of view) taken in 
1948--1996 with the 40\,cm astrograph \citep{2010ASPC..435..135S}.
The plates are being digitized with the Epson Expression~11000XL scanner at 
2400\,dpi resolution ($1.37^{\prime\prime}$/pix, 16~bit grayscale).
These digitized plates are used for variable star research 
\citep{2008AcA....58..279K,2010ARep...54.1000K,2014ARep...58..319S,2014aspl.conf...79S,2016arXiv160503571S}.

\section{Data processing algorithm}

Conventional software used for CCD photometry is often not applicable to
digitized photographic plates as it relies on the assumptions that the image
detector responds linearly to the number of incoming photons and that the
image astrometric solution may be well represented with the polynomial
distortion corrections. The \texttt{VaST} software \citep{2017arXiv170207715S}
that we develop, relies on the commonly used
\texttt{SExtractor} \citep{1996A&AS..117..393B}, \texttt{Astrometry.net}
\citep{2010AJ....139.1782L,2008ASPC..394...27H} 
and \texttt{WCSTools} 
\citep{2014ASPC..485..231M}
and combines them with the original code developed to carefully circumvent the above limitations. 
Here we summarize the key design points. 

{\underline{\it Astrometry}}
Flatbed scanners introduce systematic errors in source positions. 
After computing with \texttt{Astrometry.net} an approximate plate solution, 
for each detected source we use nearby UCAC4 \citep{2013AJ....145...44Z} stars 
to compute the {\it local correction} to the source position. The resulting accuracy of
$<1^{\prime\prime}$ is 
sufficient to match the imaged sources to
USNO-B1.0 \citep{2003AJ....125..984M} (the ``A'' series plates go deeper than UCAC4).

\begin{figure*}[!htb]
 \centering
 \includegraphics[width=0.48\textwidth]{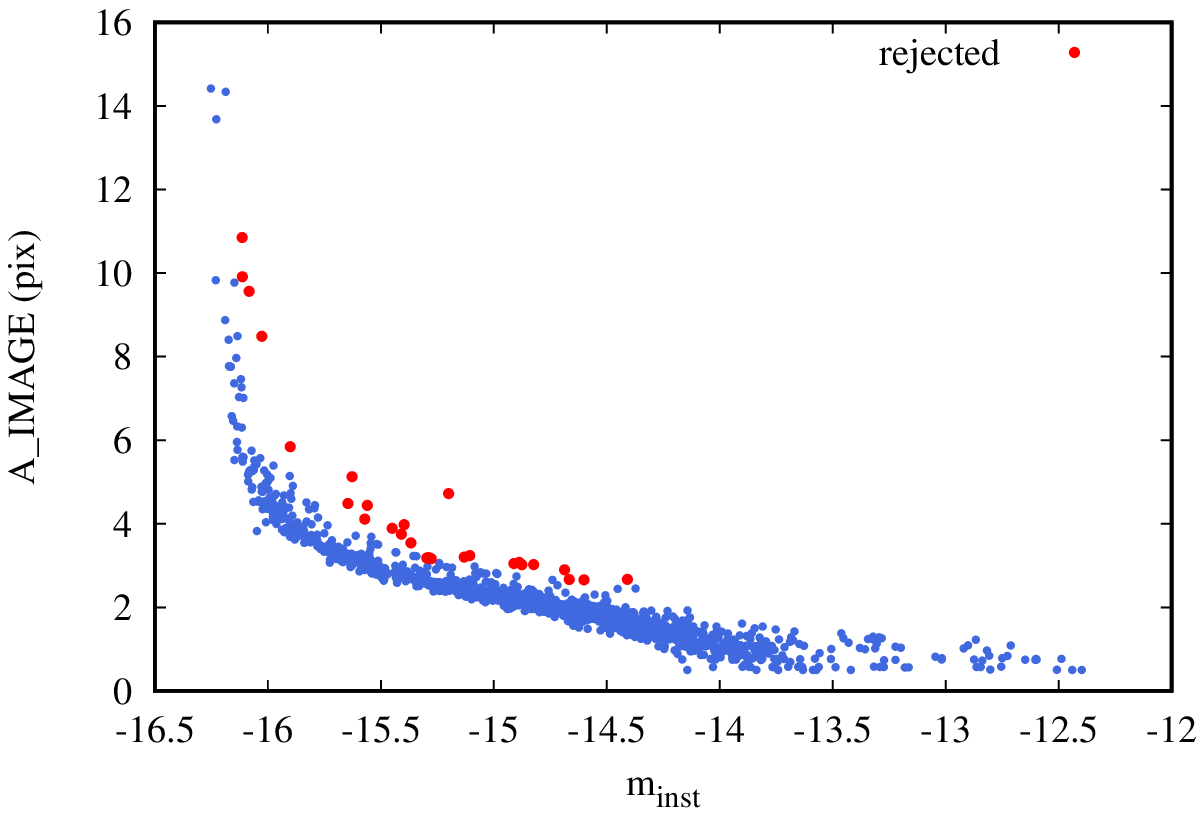}~~
 \includegraphics[width=0.48\textwidth]{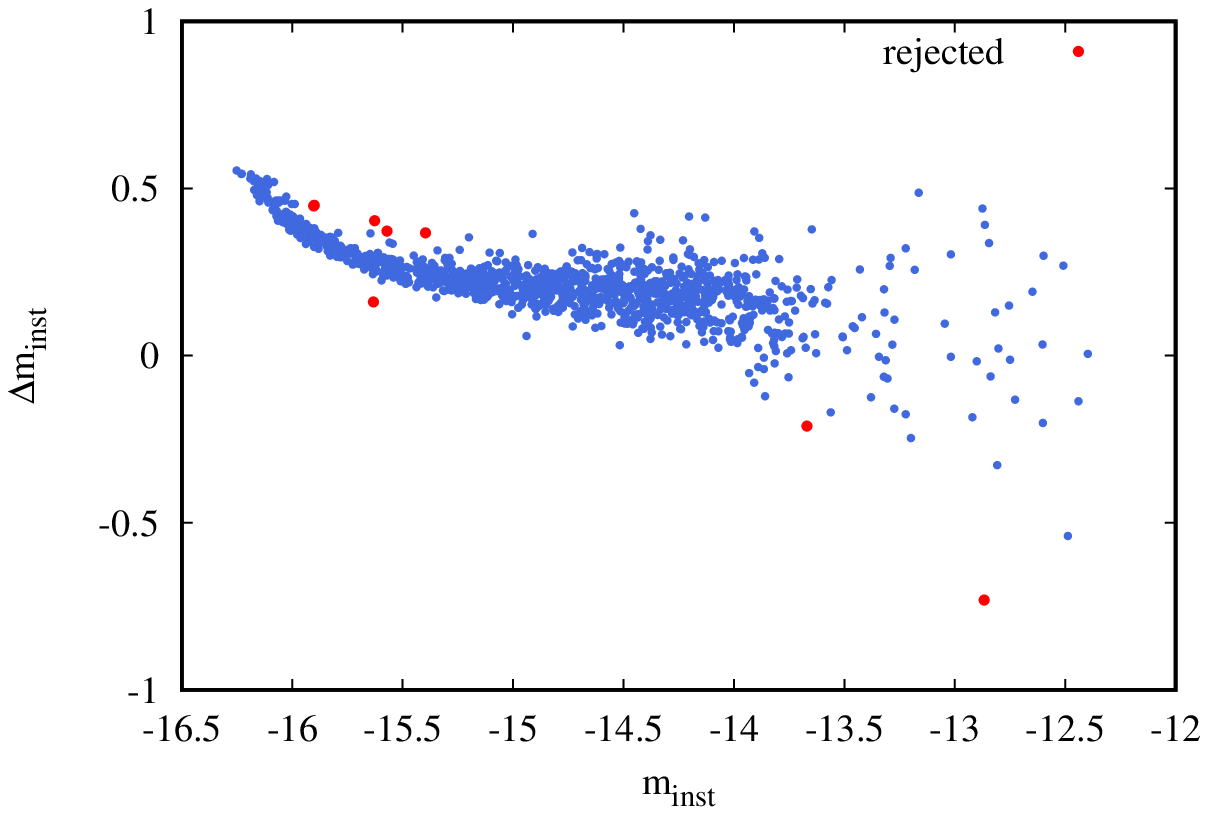}
 \caption{Magnitude vs. source size along its major axis {\it (left)} and magnitude vs. difference between the magnitudes
measured in two concentric circular apertures {\it (right)}. One aperture is
30\,\% larger than the other. The identified blended objects are highlighted. 
This blend rejection procedure is applied automatically to each image.}
 \label{fig:magsize}
\end{figure*}

{\underline{\it Filtering}}
The brightness of blended stars cannot be accurately measured with aperture
photometry. We do not employ image subtraction or PSF-fitting 
considering the non-linear response of the photographic image detector 
\citep[but see][]{2016arXiv161000265S}, so we need a way to discard the
unreliable measurements.
Two types of plots are used to identify 
blended sources as they appear as 
outliers in ``magnitude vs. source size'' and 
``magnitude vs. magnitude difference between two concentric apertures'' plots
(Fig.~\ref{fig:magsize}).

{\underline{\it Photometry}}
\texttt{SExtractor} operating in the CCD mode is performing source detection and photometry 
using a fixed circular aperture. The photographic density is a non-linear function 
of the number of incoming photons. We use the function suggested by
\cite{2005MNRAS.362..542B} to approximate the relation between the instrumental 
photographic magnitude 
and the APASS B magnitude. 
The plate images are split into
overlapping $1.2\times1.2$\,deg subfields that are calibrated independently
of each other. With the ``A'' series plates the highest accuracy
relative photometry ($\sigma \approx 0.08$\,m) is reached for sources in the
magnitude range $14<B<15$. Photometric accuracy for the brighter
stars is deteriorated as they spill over the fixed-radius aperture.

\begin{figure*}[!htb]
 \centering
 \includegraphics[width=0.53\textwidth]{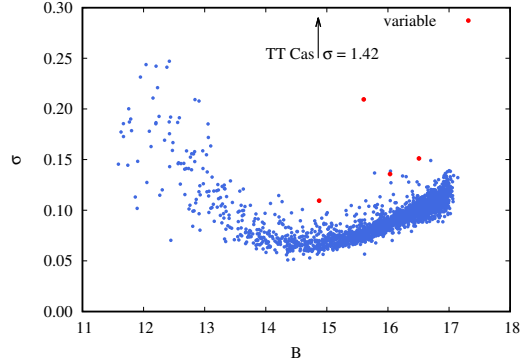}
 \caption{The magnitude vs. standard deviation ($\sigma$) 
plot highlighting
variable objects as the ones having lightcurve scatter 
larger than
most stars of similar brightness in this field. }
 \label{fig:magsigma}
\end{figure*}

\begin{figure*}[!htb]
 \centering
 \includegraphics[width=0.45\textwidth]{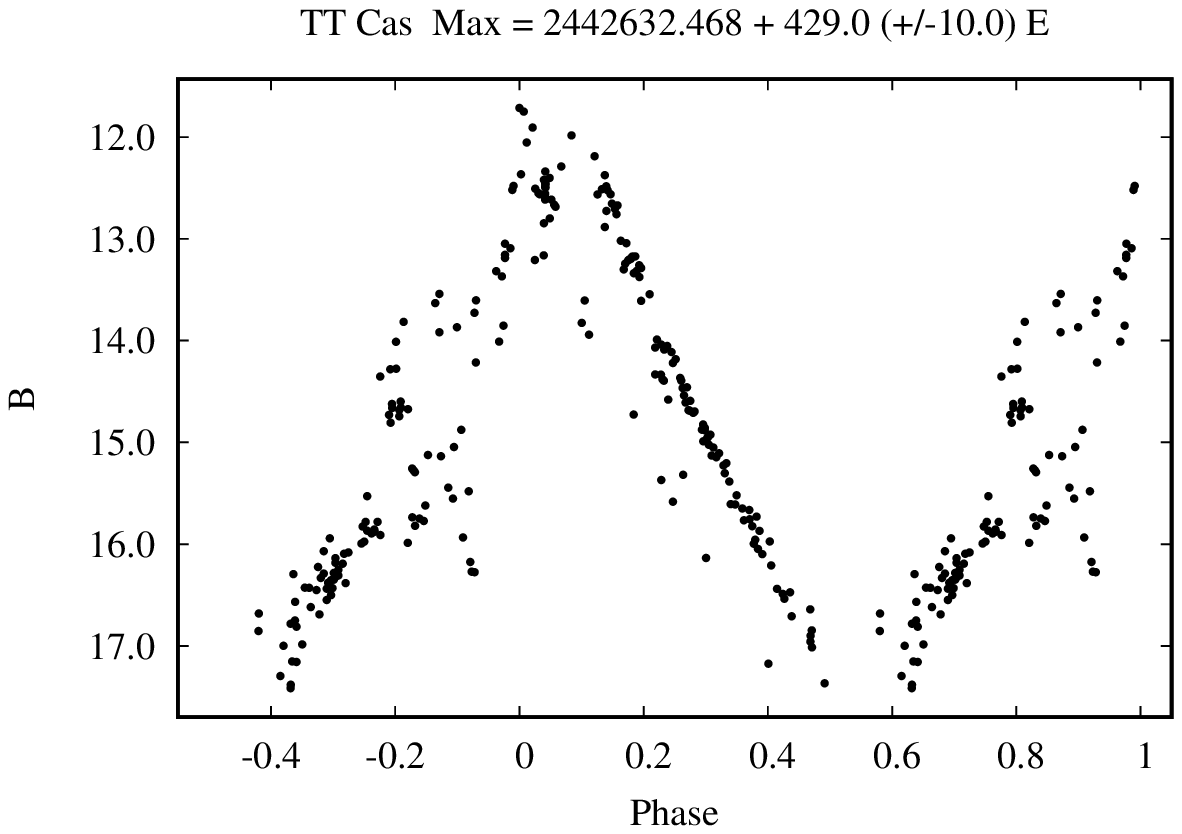}~~
 \includegraphics[width=0.25\textwidth,clip=false,trim=0cm -1.5cm 0cm 0cm]{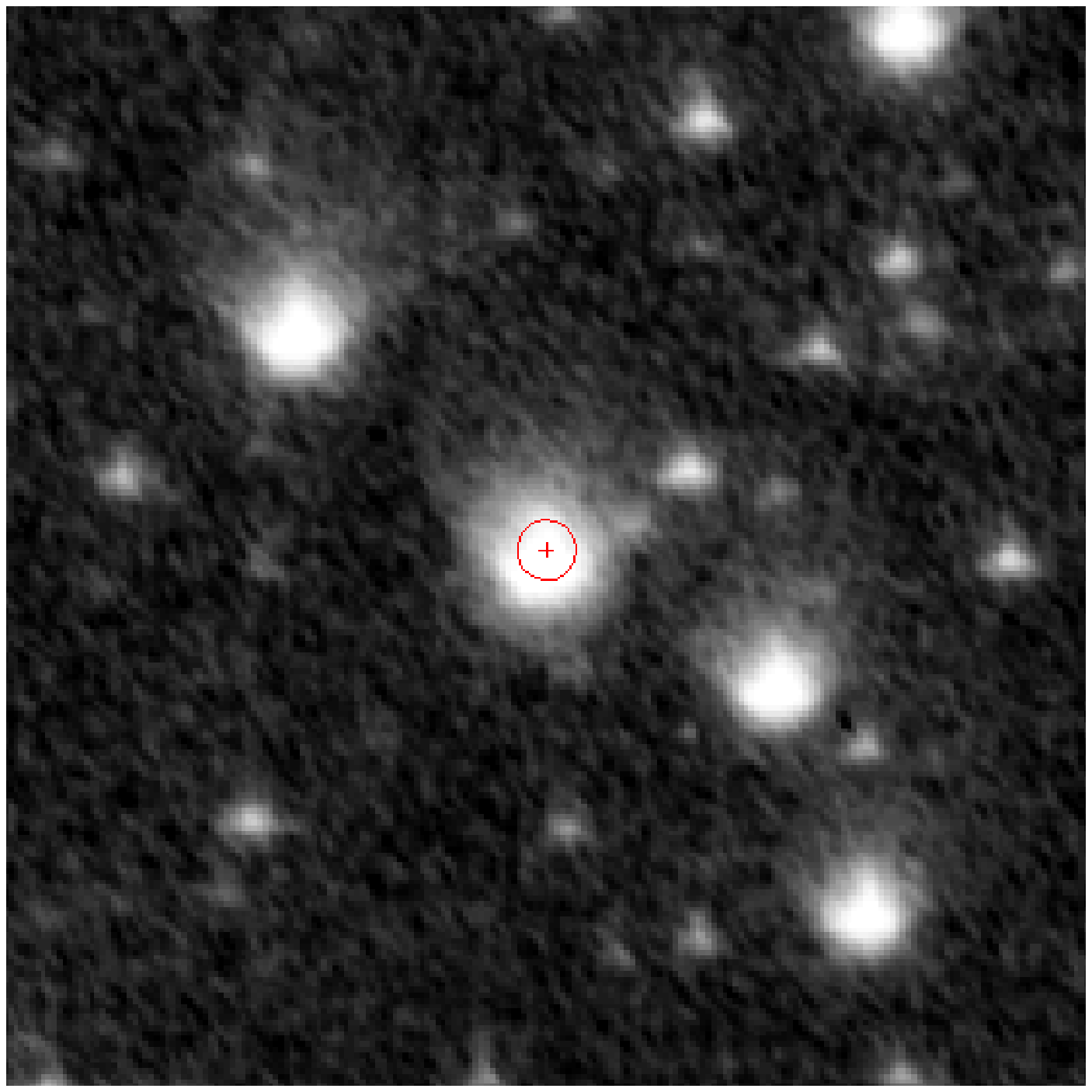}
 \includegraphics[width=0.25\textwidth,clip=false,trim=0cm -1.5cm 0cm 0cm]{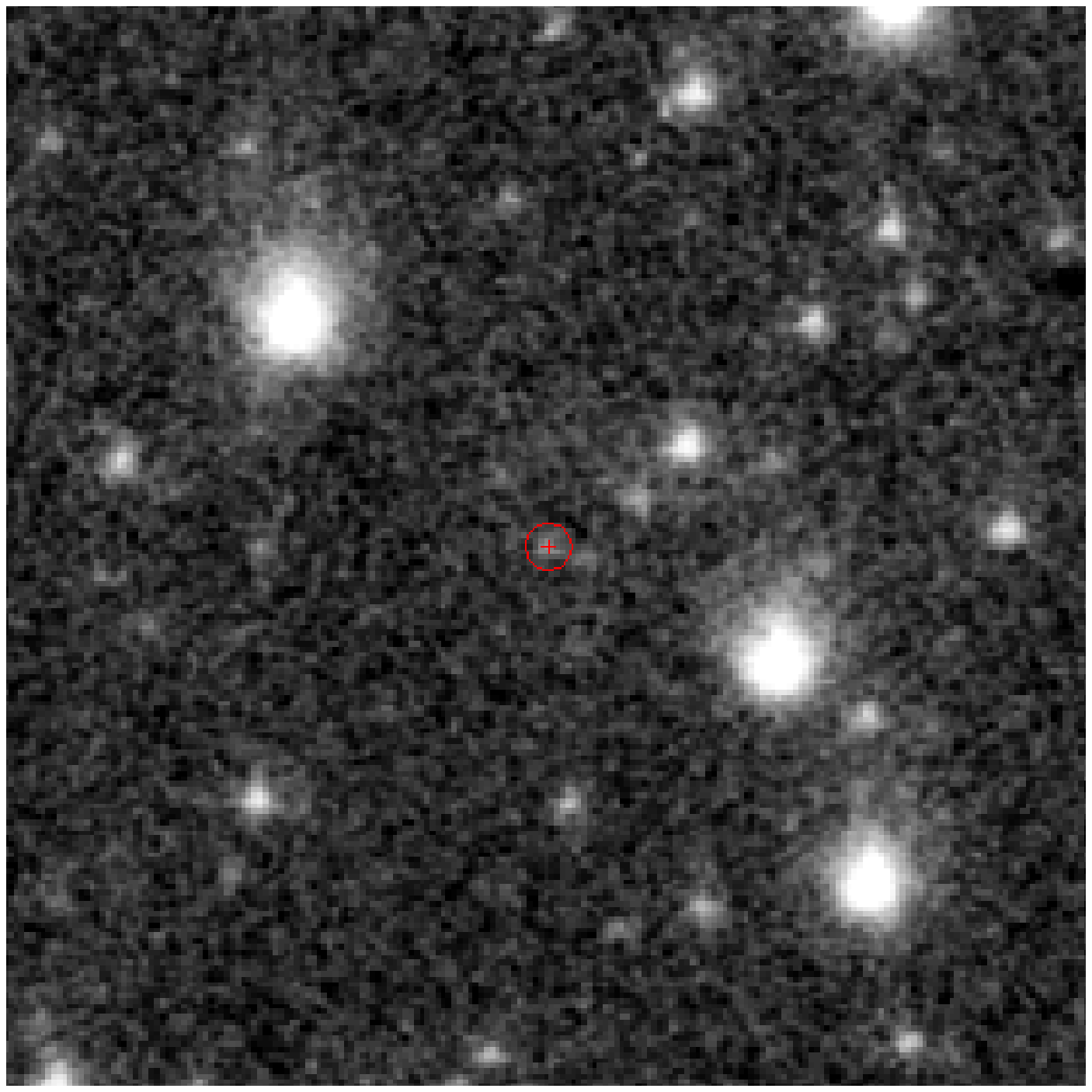}
 \caption{{\it (left)} Phased photographic lightcurve of the Mira-type
variable TT~Cas. The lightcurve includes 238 measurements
obtained over 25 years. The derived period of $429\pm10$\,d
is not consistent with the previously published period of 396\,d.
{\it (right)}~TT~Cas (indicated by the red marker) imaged
with the 40\,cm astrograph at maximum on 1975-08-07 ($B=11.7$)
and minimum on 1971-08-24 ($B=17.4$).}
 \label{fig:ttcas}
\end{figure*}

{\underline{\it Example results}}
Fig.~\ref{fig:magsigma} presents the magnitude--$\sigma$ plot for a typical subfield
that is about half-way between the plate center and the edge. This subfield
includes the Mira-type variable TT~Cas that has the variability amplitude
covering the full accessible magnitude range (Fig.~\ref{fig:ttcas}).

\section{Summary}

We present an overview of the data processing steps used to extract
lightcurves from a series of digitized photographic images.
The \texttt{VaST} code 
implementing these 
steps relies heavily on 
\texttt{SExtractor} and
\texttt{Astrometry.net} for source detection and astrometry. 
\texttt{VaST} takes into account the non-linear
response of the photographic emulsion and can tolerate the hacksaw-like systematic
errors in source positions introduced by the scanner.
While our tests are 
confined to the plates of one specific 
series of the Moscow collection,
the proposed processing strategy is applicable to any series of 
digitized photographic images of the sky.

\end{document}